\documentclass{article}
\usepackage{spconf,amsmath,graphicx}

\usepackage{booktabs}
\usepackage{multicol, multirow}
\usepackage{tabularx}
\usepackage{amssymb}
\usepackage{adjustbox}
\usepackage[dvipsnames]{xcolor}

\usepackage[ruled,vlined]{algorithm2e}

\title{Contrastive speech mixup for low-resource keyword spotting}
\name{\begin{tabular}{c} 
Dianwen Ng$^{1,2}$, Ruixi Zhang, Jia Qi Yip$^{1,2}$, Chong Zhang$^{1}$, Yukun Ma$^{1}$,\\
Trung Hieu Nguyen$^{1}$,
Chongjia Ni$^{1}$,
Eng Siong Chng$^{2}$, 
Bin Ma$^{1}$ \thanks{This work was supported by Alibaba Group through Alibaba Innovative
Research (AIR) Program and Alibaba-NTU Singapore
Joint Research Institute (JRI), Nanyang Technological University,
Singapore.}
\end{tabular}}

\address{$^1$Alibaba Group \\
 $^2$School of Computer Science and Engineering, Nanyang Technological University, Singapore}
 
\begin{document}
%\ninept
%
\maketitle
\begin{abstract}
Most of the existing neural-based models for keyword spotting (KWS) in smart devices require thousands of training samples to learn a decent audio representation. However, with the rising demand for smart devices to become more personalized, KWS models need to adapt quickly to smaller user samples. To tackle this challenge, we propose a contrastive speech mixup (CosMix) learning algorithm for low-resource KWS. CosMix introduces an auxiliary contrastive loss to the existing mixup augmentation technique to maximize the relative similarity between the original pre-mixed samples and the augmented samples. The goal is to inject enhancing constraints to guide the model towards simpler but richer content-based speech representations from two augmented views (i.e. noisy mixed and clean pre-mixed utterances). We conduct our experiments on the Google Speech Command dataset, where we trim the size of the training set to as small as 2.5 mins per keyword to simulate a low-resource condition. Our experimental results show a consistent improvement in the performance of multiple models, which exhibits the effectiveness of our method.
\end{abstract}
\begin{keywords}
Keyword Spotting, Data Augmentation, Contrastive Learning, Low-resource
\end{keywords}
\section{Introduction}
\label{sec:intro}
\vspace{-0.1cm}
The inbuilt voice command system found in many smart devices has brought greater convenience to our lives. For instance, we can effortlessly activate the devices with a single wake-up command like ``Hey Siri" to schedule an important meeting later in the afternoon. Obtaining such a system that recognizes these commands involves building a keyword-spotting (KWS) model to detect predetermined words in a continuous utterance. Operationally, the system converts the raw audio into temporal-spectral features before passing them to an acoustic neural net to predict the best keyword classes that minimize the error rate. 

Recently, there have been multiple successes in developing highly accurate KWS networks \cite{sainath2015convolutional, gong2021ast, ng2022small, peter2022end}. These network architectures include utilizing transformer blocks \cite{berg2021keyword, ding2022letr}, convolutional blocks \cite{ng2022convmixer, choi2019temporal}, auto-regressive layers \cite{rybakov2020streaming} and hybrid structures of these \cite{rybakov2020streaming, zeng2019effective} to create a deep acoustic neural-based model. 
% Amongst the networks, convolutional-based models have achieved the best-performing benchmark for lightweight non-streaming applications, whereas auto-regressive recurrent models have emerged as the method of choice for streaming applications \cite{lopez2021deep}.
Most of these neural-based models face the challenge of having inadequate annotated training data. Learning a decent audio representation for the pre-defined keywords usually requires thousands of training samples to avoid over-fitting \cite{chen2022noise}, especially for deeper networks. Nevertheless, with the rising demand for personalized smart devices, there is a need for customized KWS systems to adapt quickly with limited user samples.
  
To overcome this challenge, researchers have explored many data augmentation techniques, such as mixing tiny noise distortion, time shift, time stretch and SpecAugment \cite{wei2020comparison, park2019specaugment}, to improve the generalizability of deep neural networks under low-resource conditions. These augmentations inject small variability to the data instances and prevent the model from memorizing the dataset, thus reducing the tendency of overfitting. However, there are limited types of perturbation available in speech preprocessing, which restrain the diversity of the data augmentations.  

In this paper, we attempt to improve the robustness of KWS model under low-resource conditions. Specifically, we aim to train a deep network that achieves better performance on very small training sets (i.e., 2.5~mins, 5~mins, 10~mins). To achieve this, we propose a contrastive speech mixup (CosMix) learning algorithm for low-resource KWS, which is inspired from the input mixup algorithm \cite{zhang2018mixup}. The mixup algorithm is a regularization technique that fits the network with two linearly interpolated samples to their corresponding soft labels. Although the vanilla method has been demonstrated to be effective in several works, \cite{yun2019cutmix, kim2020co} have shown that mixing samples sometimes produce spatially ambiguous and unnatural instances which confuses the model, especially when the model attempts to locate non-existent spatial information. Hence, we introduce an auxiliary contrastive loss that imposes an additional constraint to maximize the relative similarity of the original individual samples to the augmented version (which we denote as positive pairs). With the contrastive loss, the pre-mixed sample representation provides supervision to the model to pull the agreeing attributes between the positive mixed-individual samples. This supervision reduces ambiguity by providing awareness of the mixing utterances. Furthermore, it also offers some sense of enhancement, similar to \cite{ng2022i2cr, hu2022interactive}, to generate content-rich speech encoding. 
% Besides, enhancing two mixed audio waveforms drives separation by fostering minimal (i.e. less complexity) and sufficient (i.e. better fidelity) model encodings. Consequently, the model learns to develop more efficient representations and attain better generalizability, especially for small-data training. 
Our experimental results show that CosMix obtains a consistent performance improvement over the vanilla mixup augmentation over various KWS models on variant training data sizes. Among the compared models, Keyword ConvMixer performs the best with 90\% accuracy on 5\% (2.5 mins per keyword) of the training data. The overall results exhibit the effectiveness of our proposed CosMix learning algorithm despite the simplicity of its design and incurring negligible computational overhead.
\vspace{-0.3cm}
\section{Related Work}
\vspace{-0.1cm}
Multiple attempts have been made to improve the vanilla mixup algorithm in different fields. In image classification, Manifold Mixup \cite{verma2019manifold} leverages semantic interpolations as an additional training signal to achieve smoother decision boundaries at multiple levels of representation. CutMix \cite{yun2019cutmix} replaces a region of the image with a patch sampled from another training image to preserve the naturalness of training images and enhances the model's robustness against input corruptions and its out-of-distribution detection performances. In speech signal processing, MixSpeech \cite{meng2021mixspeech} shares similarities with Mixup, mixes two input speech signal sequences and combines two loss functions regarding the text output to facilitate the application of the mixing algorithm in sequential tasks such as ASR. Moreover, L-mix  \cite{kang2022mix} utilizes the instance mix (i-mix) regularization for training a self-supervised speaker embedding system to improve the training stability and speaker verification performance. In contrast, our work focuses on a different variant which we have shown to be effective for low-resource keyword spotting.

\section{Methodology}
\label{sec:method}
% \begin{figure*}[t]
%   \centering
%   \includegraphics[width=0.75\linewidth, height=5cm]{model1.png}
%   \vspace{-0.3cm}
%   \caption{An illustration of the model architecture for contrastive speech mixup (CosMix). The proposed approach includes audio mixup augmentation and contrastive learning for the mixup instances, where \textit{Project} refers to the projector that maps the latent vector embeddings to the projected dimension of 128, and \textit{sg} denotes the stop-gradient function.}
%   \label{fig:model}
% \vspace{-0.1cm}
% \end{figure*}

\begin{figure}[t]
  \centering
  \includegraphics[width=1\linewidth, height=4cm]{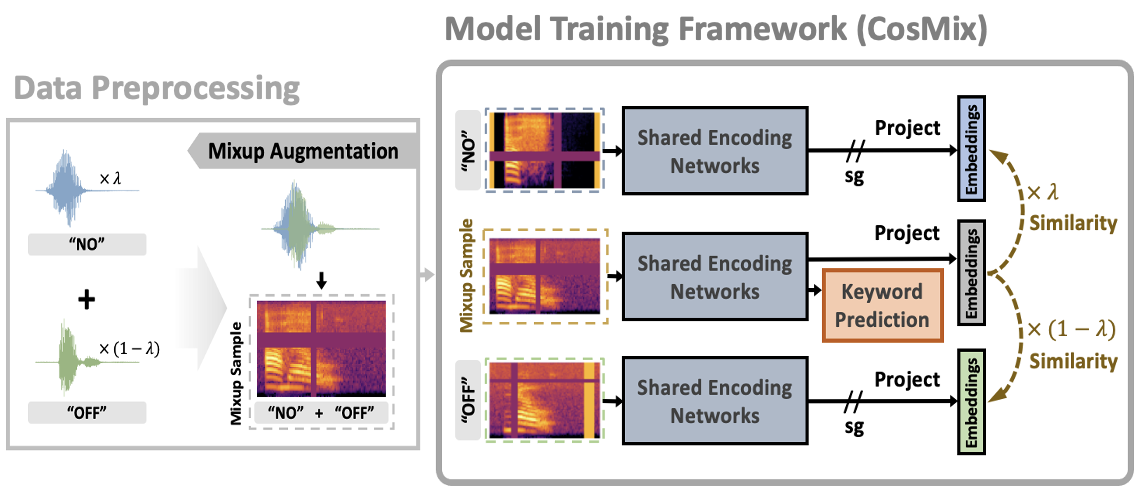}
  \vspace{-0.6cm}
  \caption{An illustration of the model architecture for contrastive speech mixup (CosMix). The proposed approach includes audio mixup augmentation and contrastive learning for the mixup instances, where \textit{Project} refers to the projector that maps the latent vector embeddings to the projected dimension of 128, and \textit{sg} denotes the stop-gradient function.}
  \label{fig:model}
\vspace{-0.15cm}
\end{figure}

\subsection{Mixup Augmentation}
The vanilla mixup augmentation employs the principle of vicinal risk minimization to encourage the classifier to behave linearly within training examples. This attribute reduces undesirable variability when performing inferences on unseen instances. During data pre-processing, we draw two audio instances at random from the training set to construct a virtual training example by
\begin{equation}
\label{eq:1}
\begin{split}
    \tilde{x} &= \lambda x_i + (1-\lambda)x_j \\
    \tilde{y} &= \lambda y_i + (1-\lambda)y_j 
\end{split}
\end{equation} where $i$, $j \in {1,..N}$ denotes the indices of the training set, $x$ and $y$ represent the raw input waveform and one-hot label encodings, respectively. $\lambda \sim \text{Beta}(\alpha, \alpha)$, for $\alpha \in (0, \infty)$ is the interpolating parameter that determines the amount of content to be linearly mixed. Given the virtual input-label pairs $(\tilde{X}, \tilde{y})$, with $\tilde{X}$ being the STFT temporal-spectral features of $\tilde{x}$, we compute the loss as follows
\begin{equation}
    \mathcal{L}_{\text{mix}} = \lambda\cdot \text{CE}(f(\tilde{X}), y_i) + (1 - \lambda)\cdot \text{CE}(f(\tilde{X}),y_j)
\end{equation} where $f(.)$ is an acoustic encoding model, and CE refers to the standard cross-entropy loss.
   
\subsection{The CosMix Learning Algorithm}
Despite succeeding in a wide range of speech applications, the mixup augmentation may produce highly distorted signals from two overlapping speeches that are not natural or unfavorable for the task. The noisy input may create confusion that causes peculiar spikes in the model errors, which may dominate the effective gradient and hurt the network convergence \cite{yun2019cutmix, kim2020co}. To tackle this problem, we introduce an auxiliary component that uses contrastive learning to maximize the relative similarity between the original mixing samples and the augmented samples. Our method is inspired by \cite{ng2022i2cr} which uses the agreeing attributes between two augmented (positive) views to enhance content-rich speech encoding. In this case, we make use of two mixed utterances with the aforementioned training method which fosters minimal (i.e. less complexity) and sufficient (i.e. higher fidelity) embeddings. This drives the model to generate more effective representations and attain greater generalizability under low-resource conditions.

Specifically, the training procedure utilizes parallel samples of two utterances $X_i$, {$X_j$} and their mixup $\tilde{X}$. %We first convert all three into time-frequency features using STFT. 
We apply random augmentations that perform independently on three parallel samples. Subsequently, we pass them into a shared encoding network where the bottleneck representations will be projected and used in contrasting with the mixup and pre-mix individual pair. At the same time, only the mixup representations are sent for keyword prediction. An illustration of the training framework is presented in Fig. \ref{fig:model}. Then, to compute the loss for contrastive learning, we perform L2-norm on all projected embeddings before utilizing the mean square error to measure the similarity between the normalized projections. We define the loss by 
\begin{equation}
    \mathcal{L}_\text{cos}(\tilde{X}, X_r) = - \frac{\langle f_p(\tilde{X}), f_p(X_r)\rangle}{\|f_p(\tilde{X})\|_2 \| f_p(X_r)\|_2}
\end{equation} where $f_p(.)$ is the projector of the model and $r\in \{i, j\}$. %refers to the two pre-mixed utterances. 

Additionally, we implemented a mixup ratio of 50\% in this work, where there is a half chance that the model is learning without mixup. Nevertheless, we included contrastive loss with its augmented views to reduce variability with perturbed samples to achieve less complex embeddings. Therefore, the relative contrastive loss is weighted as follows
\begin{equation}
  \Lambda_r =
    \begin{cases}
      \lambda, & \text{if $r=i$ \& $r\neq j$}\\
      1 - \lambda, & \text{if $r=j$ \& $r\neq i$}\\
      1, & \text{if $r=i=j$}
    \end{cases}       
\end{equation} and the complete training loss is given by
\begin{equation}
    \mathcal{L} = \mathcal{L}_\text{mix} + \beta \sum_{r\in \{i,j\}}\Big(\Lambda_r \cdot \mathcal{L}_\text{cos}(\tilde{X}, X_r)\Big) 
\end{equation} where %$X_r$, $r\in \{ i,j\} $ are STFT features of the original utterances $x_i$ and $x_j$. 
$\beta$ is the penalizing parameter that weighs the contribution of the contrastive loss, which is set to $0.5$ in this work. 

\vspace{-0.1cm}
\section{Experiment}
\label{sec:experiment}
\vspace{-0.4cm}
\subsection{Dataset}
In this work we utilize the Google Speech Command V2 dataset \cite{warden2018speech}. The dataset contains a total of 105,000 utterances with 35 unique words. Each audio sample is stored as a one-second (or less) wav format file sampling at 16kHz. To ensure the reproducibility of our code, we employ the official train, validation, and test split provided for the subset of 10 keyword classes, which covers the words: ``up", ``down", ``left", ``right", ``yes", ``no", ``on", ``off", ``go" and ``stop". To simulate low-resource KWS conditions, we partition the utterances according to the speaker for each word and trim the size of the corresponding training sets. In our case, we experiment with 5\%, 10\%, 20\%, 30\%, and 50\% of the train set. The trimming corresponds to an average of 2.5~mins, 5~mins, 10 mins, 15~mins, and 25~mins training data for each word. By trimming the speaker partition, we reduce the diversity of our training data and increase the learning difficulty to adapt to a wider population in actual deployment. This size of training data approaches what would be available for training a personalized KWS model in real applications.

\subsection{Experimental Setup}
\noindent\textbf{Input Feature} - We convert all wav files to a 64-dimensional log Mel filterbank (FBank) with a window size of 25ms and 10ms shift. We fixed the resolution of our FBank at 98 $\times$ 64 (i.e. equivalent to 1s of the utterance). Commands shorter than 1s will be zero-padded to the right. During training, we augment our samples, which includes random time shifting in the range of -100 to 100ms, and random time stretching \cite{ko2015audio} between the factor of 0.9 to 1.1. Furthermore, we apply SpecAugment with the masking size for time and spectral of 13 and 7, respectively. Lastly, the mixup ratio for our main experiment in Table \ref{tbl:main} is 0.5 with a \textit{Beta($10, 10$)} distribution. This applies to the vanilla mixup and CosMix. 

\noindent\textbf{Training Details} - In our experiment, we picked two popular categories of neural-based models that are frequently used in the industry, namely, the transformer-based (i.e. KWT-1 and KWT-3 \cite{berg2021keyword}) and convolutional-based (i.e. Keyword ConvMixer \cite{ng2022convmixer} and ResNet18 \cite{he2016deep}) KWS networks. These models represent the recent state-of-the-art for different conditioning environments with various model sizes and complexity. We observe that the most lightweight model, ConvMixer \cite{ng2022convmixer}, consists of only 0.1M parameters. Additionally, we add a projector to every model that maps the latent vector embeddings to a projected dimension of size 128. The projector consists of a linear dense block with ReLU activation. All models are trained with a batch size of 128. The initial learning is 5e-3, and we employ a step decay of rate 0.85 every four epochs from the 5th to 70th epoch. Lastly, we use the Adam optimizer and binary cross-entropy loss in the optimization.

\begin{figure}
    \centering
    \includegraphics[width=0.8\linewidth, height=5.5cm]{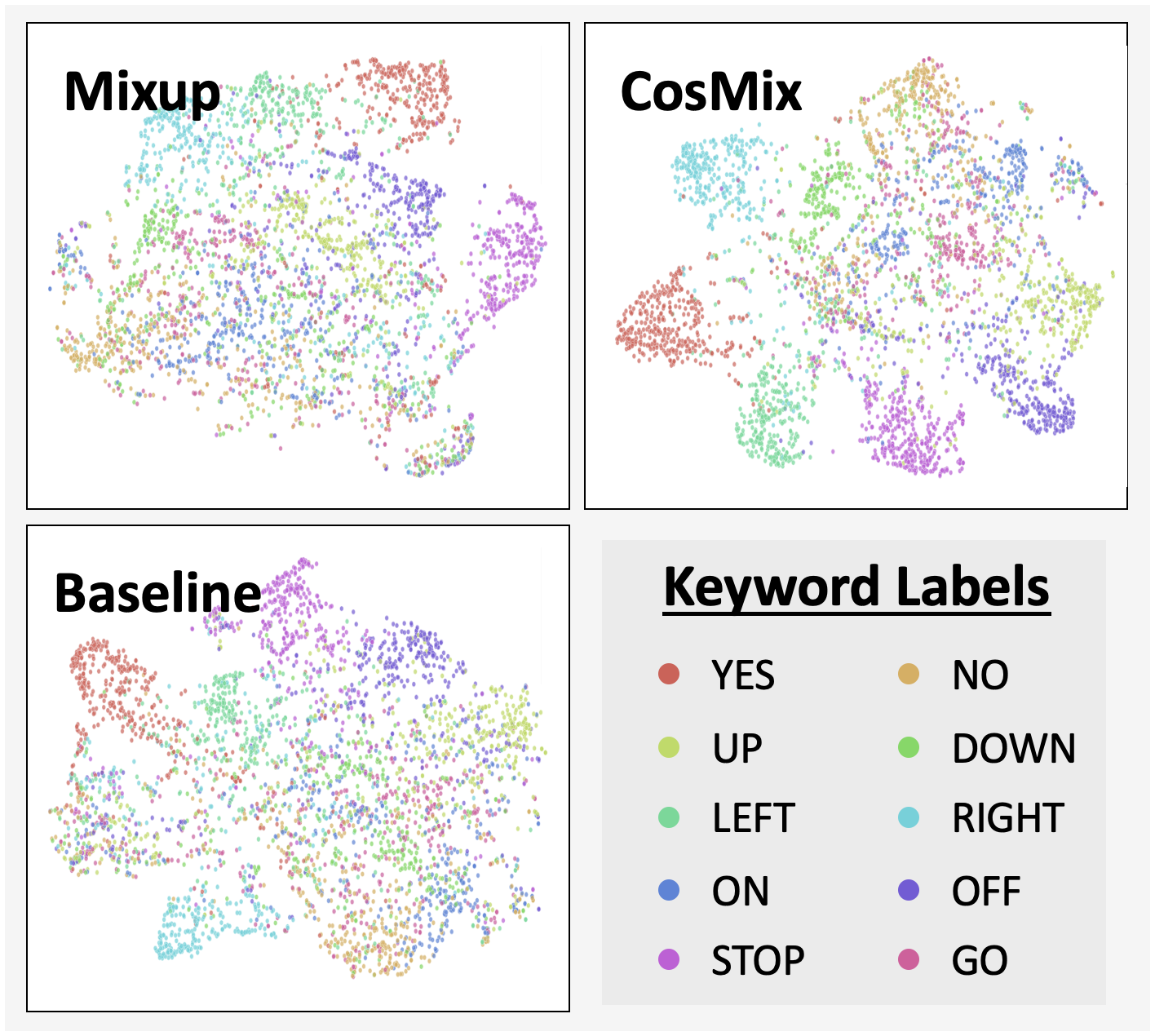}
    \vspace{-0.3cm}
    \caption{t-SNE plots that compare the embeddings on different techniques with 20\% train set of KWT-3 (after 200 epochs).}
    \label{fig:tsne}
    \vspace{-0.3cm}
\end{figure}

\vspace{-0.2cm}
\section{Experimental Results}
\label{sec:result}

% \usepackage{multirow}
\begin{table*}[!t]
\centering\small
\renewcommand{\arraystretch}{1.1}
\tabcolsep=0.29cm
\caption{Experimental results on Google Speech Command-V2 (10 classes) testing dataset. We measure the accuracy of the keyword classification on the official test set over the different sizes (i.e. 5\%, 10\%, 20\%, 30\%, 50\%) of training data.}
\vspace{0.1cm}
\label{tbl:main}
\begin{tabular}{l|c|c|ccccc}
\hline
\multirow{3}{*}{Model} & \multirow{3}{*}{Model Size (M)} & \multirow{3}{*}{Augmentation} & \multicolumn{5}{c}{Size of Training Data (\% of Train Set)} \\ \cline{4-8} 
 &  &  & \multicolumn{1}{c}{5\%} & \multicolumn{1}{c}{10\%} & \multicolumn{1}{c}{20\%} & \multicolumn{1}{c}{30\%} & \multicolumn{1}{c}{50\%} \\
 &  &  & \multicolumn{1}{c}{(2.5 mins)} & \multicolumn{1}{c}{(5 mins)} & \multicolumn{1}{c}{(10 mins)} & \multicolumn{1}{c}{(15 mins)} & \multicolumn{1}{c}{(25 mins)} \\ \hline

\multirow{3}{*}{ResNet18 \cite{he2016deep}} & \multirow{3}{*}{11.9} & Baseline & 0.828 & 0.902 & 0.934 & 0.963 & 0.971 \\
 &  & Vanilla Mixup & 0.822 & 0.909 & 0.947 & 0.968 & 0.978 \\
 &  & CosMix (Ours) & \textbf{0.841} & \textbf{0.919} & \textbf{0.955} & \textbf{0.970} & \textbf{0.981} \\ \hline
 
\multirow{3}{*}{KWT-3 \cite{berg2021keyword}} & \multirow{3}{*}{5.4} & Baseline & 0.465 & 0.613 & 0.635 & 0.677 & 0.732 \\
 &  & Vanilla Mixup & 0.524 & 0.607 & 0.643 & 0.715 & 0.766 \\
 &  & CosMix (Ours) & \textbf{0.566} & \textbf{0.687} & \textbf{0.765} & \textbf{0.804} & \textbf{0.841} \\ \hline
 
\multirow{3}{*}{KWT-1 \cite{berg2021keyword}} & \multirow{3}{*}{0.6} & Baseline & 0.683 & 0.836 & 0.882 & 0.927 & 0.950 \\
 &  & Vanilla Mixup & 0.625 & 0.853 & 0.892 & 0.933 & 0.951 \\
 &  & CosMix (Ours) & \textbf{0.701} & \textbf{0.863} & \textbf{0.911} & \textbf{0.935} & \textbf{0.960} \\ \hline

\multirow{3}{*}{Keyword ConvMixer \cite{ng2022convmixer}} & \multirow{3}{*}{0.1} & Baseline & 0.887 & 0.926 & 0.954 & 0.965 & 0.971 \\
 &  & Vanilla Mixup & 0.897 & 0.939 & \textbf{0.963} & \textbf{0.973} & \textbf{0.978} \\
 &  & CosMix (Ours) & \textbf{0.902} & \textbf{0.940} & 0.962 & \textbf{0.973} & 0.976 \\ \hline

\end{tabular}
\vspace{-0.4cm}
\end{table*}

We use the Google Speech Command official test set (10 classes) to evaluate the model performance. The baseline model in our experiment learns without any mixup augmentation. Nevertheless, we also compare CosMix to the vanilla mixup augmentation to determine the performance gain contributed by the contrastive loss in CosMix model training. Table \ref{tbl:main} shows the accuracy of our KWS model over four different architectural builds. We observe that all models suffer performance degradation when training on a small dataset. The KWT-3 sees the largest performance degradation, achieving an accuracy of only 46.5\% with baseline training at 5\% train data. Nevertheless, CosMix has generally achieved the best performance over various training set sizes regardless of the model used. Moreover, the performance gain is larger with less training data (5\% size), where the relative increase in performance is as high as 21.7\% for KWT-3. Studying the individual models, we notice that the transformer-based models are more vulnerable to performance degradation in the low-resource setting. This could be due to the lack of training samples to learn the highly complex attention mechanism. Even in this case, CosMix helped to alleviate this problem with better regularization. Finally, the convolutional-based models perform the best, with Keyword ConvMixer attaining the highest score of 90\% accuracy on the 5\% training set using the CosMix. 

To investigate the quality of the acoustic representations of different techniques, we visualize the embeddings using t-SNE plots in Fig. \ref{fig:tsne} with KWT-3 training on 20\% train data. The baseline setup managed to differentiate ``right" and ``yes" from the other commands. However, the model failed to do so for the rest. The clusters become slightly spaced as embeddings improved with mixup augmentation. The clusters in CosMix are the most separable, with only short consonant
words are tied together, which demonstrates our approach to be effective in learning precise and content-rich representations. 

% \begin{figure}
%     \centering
%     \includegraphics[width=0.85\linewidth, height=6.cm]{ICASSP/tsne.png}
%     \caption{t-SNE plots that compare the latent embeddings of different augmentation technique.}
%     \label{fig:tsne}
%     \vspace{-0.2cm}
% \end{figure}
\vspace{-0.3cm}
\subsection{Ablation Study}
\vspace{-0.1cm}
In this section, we investigate the impact of the performance using different parameters on the two mixup algorithms. In particular, we look at changing the mixup ratio and the interpolating weights derived from the beta distribution. When $\alpha$ in the beta distribution, $\textit{Beta}(\alpha, \alpha)$, is less than 1, we obtained a convex shape curve where the amount of audio mixing tends to dominate on one side. However, when $\alpha$ is bigger than 1, the curve becomes more concave, and it is more likely that the two audio are mixed proportionally. Table \ref{tbl:ablation} presents the result of our findings. Firstly, $\textit{Beta}(10, 10)$ is generally better than $\textit{Beta}(0.5, 0.5)$, which suggests that the model can learn more effectively with an equal proportionally mixed audio sample for KWS task. Secondly, the optimal mixing ratio differs between Mixup and CosMix. Mixup has the best performance with a mixing ratio of 30\%, whereas CosMix got its best result at 50\%. Lastly, users should be careful when performing hyper-parameters tuning as we observed two peaks in the model performance. This seems to be highly influenced by the bi-modal distribution of beta.

\begin{table}[ht]
\centering\small
\renewcommand{\arraystretch}{1.1}
\tabcolsep=0.25cm
\vspace{-0.3cm}
\caption{Accuracy on official test set based on 20\% train set of ResNet18 over different mixing ratio of the mixup.}
\vspace{0.1cm}
\label{tbl:ablation}
\begin{tabular}{c|cc|cc}
\hline
\multirow{2}{*}{Mixing Ratio (\%)} & \multicolumn{2}{c|}{\textit{Beta} (0.5, 0.5)} & \multicolumn{2}{c}{\textit{Beta} (10, 10)} \\ \cline{2-5} 
 & Mixup & CosMix & Mixup & CosMix \\ \hline
10 & \textbf{0.940} & 0.932 & 0.936 & \textbf{0.946} \\
30 & \textbf{0.949} & 0.940 & \textbf{0.953} & \textbf{0.953} \\
50 & 0.941 & \textbf{0.958} & 0.947 & \textbf{0.955} \\
70 & \textbf{0.946} & 0.945 & 0.945 & \textbf{0.952} \\
100 & 0.938 & \textbf{0.949} & \textbf{0.929} & 0.924 \\ \hline
\end{tabular}
\end{table}

\vspace{-0.6cm}
\section{Conclusion}
\label{sec:conclusion}
\vspace{-0.2cm}
In this paper, we have proposed CosMix, a novel data augmentation strategy for low-resource KWS. The CosMix model training methodology makes use of contrastive loss to mitigate the unwanted side effects of a noisy training signal arising from traditional mixup training. CosMix is able to improve model performance under low-resource conditions for a variety of model sizes. Thus, CosMix is effective as a general approach to boosting performance under low-resource conditions for applications such as personalized KWS systems for smart devices.

% Authors must proofread their PDF file prior to submission to ensure it is correct. Authors should not rely on proofreading the Word file. Please proofread the PDF file before it is submitted.

\vfill\pagebreak

\ninept
\bibliographystyle{IEEEbib}
\bibliography{refs}

\end{document}